\documentclass[conference]{IEEEtran}
\IEEEoverridecommandlockouts
\usepackage{cite}
\usepackage{amsmath,amssymb,amsfonts}
\usepackage{algorithmic}
\usepackage{graphicx}
\usepackage{textcomp}
\usepackage{xcolor}
\usepackage{comment}
\usepackage{url}
\usepackage{multirow}
\usepackage{booktabs}
\usepackage{enumitem}
\usepackage[hidelinks]{hyperref}
\usepackage[table]{xcolor}
\usepackage{amsmath}
\usepackage{makecell}
\def\BibTeX{{\rm B\kern-.05em{\sc i\kern-.025em b}\kern-.08em
    T\kern-.1667em\lower.7ex\hbox{E}\kern-.125emX}}
\begin{document}

\title{How Should We Evaluate Naturalness for \\Synthetic Speech with a Foreign Accent?
}
\author{\IEEEauthorblockN{Kentaro Onda\IEEEauthorrefmark{1}\IEEEauthorrefmark{2}, Satoru Fukayama\IEEEauthorrefmark{2}, Daisuke Saito\IEEEauthorrefmark{1}, Nobuaki Minematsu\IEEEauthorrefmark{1}}
\IEEEauthorblockA{
\IEEEauthorrefmark{1}\textit{The University of Tokyo, Japan}
\IEEEauthorrefmark{2}\textit{National Institute of Advanced Industrial Science and Technology (AIST), Japan}\\
\{ondakentaro, dsk\_saito, mine\}@gavo.t.u-tokyo.ac.jp, s.fukayama@aist.go.jp
}
}

\maketitle

\begin{abstract}
Recent advances in speech synthesis technology have enabled the generation of speech with diverse characteristics. One such characteristic is a foreign accent. In our preliminary experiments, we found that UTMOS, a widely used automatic speech quality assessment system, tended to assign lower scores to speech with stronger foreign accents, even though it is produced by a human speaker. This observation raised a question: How do humans evaluate the naturalness of foreign-accented speech? We conducted a large-scale subjective evaluation involving both native and non-native listeners under different evaluation instructions. The results showed that humans also considered accentedness when evaluating naturalness, but that defining naturalness in terms of human-likeness and selecting raters who have greater exposure to the target accent may enable more appropriate evaluation. This study aims to provide a starting point for future discussions on fairer and more appropriate evaluation of synthetic speech with a foreign accent.
\end{abstract} 


\section{Introduction}
\label{sec:intro}
Recent advances in speech synthesis technology have made it possible to generate highly natural speech in tasks such as text-to-speech (TTS) and voice conversion (VC) \cite{tan2021survey, xie-etal-2025-towards}. In particular, the synthesis of neutral reading-style speech has reached a level where it is almost perceptually indistinguishable from human speech \cite{tan2024naturalspeech, chen2024vall}. Consequently, the focus of recent research has been shifting toward the synthesis of non-neutral speech that incorporates elements such as emotion \cite{guo2023emodiff, diatlova23_ssw, yang2025emovoice}, speaking style \cite{guo2023prompttts, yang2024instructtts, ji-etal-2025-controlspeech}, and accent \cite{onda24_interspeech, onda25b_interspeech, inoue2025macst, lertpetchpun2026accent, lertpetchpun2026learning}.

The evaluation of such speech synthesis models is typically conducted through human subjective listening tests, such as mean opinion score (MOS) tests \cite{itu1996p800}. However, due to the high cost associated with these evaluations, many studies have explored automatic prediction of subjective evaluation scores \cite{tseng21b_interspeech, cooper2022sslmos, huang22f_interspeech,cooper2023voicemos, huang2024voicemos, cooper2024review}. 
In particular, UTMOS \cite{saeki22c_interspeech} has been widely used as an evaluation metric in many speech synthesis studies, and has also been adopted as an official evaluation metric in several challenges, such as the Discrete Speech Unit Challenge \cite{chang24b_interspeech } and the URGENT Challenge \cite{saijo25_interspeech, li2026icassp}.

\begin{figure}[t]
  \centering
  \vspace*{-0.5mm}
  \includegraphics[width=\linewidth]{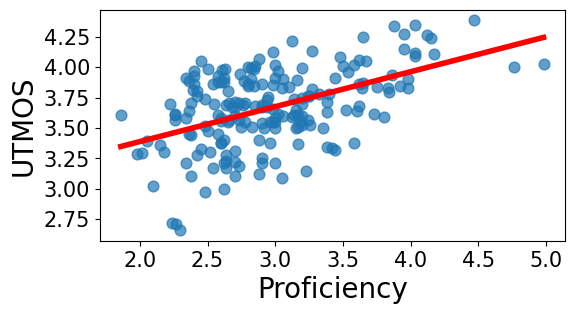}
  \vspace*{-9.5mm}
  \caption{Relationship between the proficiency of Japanese speakers of English and the naturalness scores predicted by UTMOS\cite{saeki22c_interspeech} from their utterances: speakers with lower proficiency tend to have lower naturalness scores.}
  \label{fig:erj-utmos}
  \vspace*{-5mm}
\end{figure}

Following these trends, MOS prediction systems are increasingly used even for synthesized speech with diverse characteristics beyond neutral speech. However, it has been reported that UTMOS tend to assign lower scores to certain types of speech, such as laughing speech and whispering speech \cite{nakata2024utduss}. Among these variations, we focus on speech with foreign accents. A foreign accent arises when a person speaks a non-native language, reflecting the influence of their native language. Even among speakers with the same native language, the degree of accent varies depending on their proficiency in the second language \cite{flege1995second}. 
In our preliminary experiment, we observed a clear relationship between the degree of accent and naturalness scores predicted by UTMOS. 
We used the ERJ corpus \cite{Minematsu2004DevelopmentOE}, which contains Japanese-accented English with proficiency annotations by native English speakers. 
For each speaker, we computed the average UTMOS score and compared it with the speaker's proficiency score. As shown in Fig. \ref{fig:erj-utmos}, speakers with lower proficiency tend to receive lower predicted naturalness scores, even though all samples are produced by real human speakers.

This observation raises an important concern regarding the evaluation of modern speech synthesis systems. In recent studies on accented speech synthesis, the presence of a foreign accent is an intended characteristic rather than technical defects \cite{onda24_interspeech, onda25b_interspeech, inoue2025macst, lertpetchpun2026accent, lertpetchpun2026learning}. Therefore, if evaluation systems systematically assign lower scores to speech with stronger foreign accents, they may fail to appropriately assess the quality of such systems, particularly when each system produces speech with a different degree of accent.
Moreover, automatic MOS predictors such as UTMOS are increasingly used not only for evaluation but also within the training pipeline. For example, they have been used for data filtering and corpus selection \cite{seki2025ttsops} and as reward models for reinforcement learning \cite{chen25b_interspeech, arora-etal-2026-optimizing}. In these applications, such predictors may implicitly favor native-accented speech even when accent itself is not an explicit optimization target. This could potentially limiting the diversity and inclusiveness of speech technologies.

To establish appropriate evaluation methodologies for accented speech synthesis systems, it is necessary to understand how foreign accents affect both automatic and human naturalness evaluations. It remains unclear, however, whether the observed tendency reflects a limitation specific to automatic MOS prediction systems or a broader characteristic of naturalness evaluation, including human judgments. The definition of naturalness has long been recognized as ambiguous in the speech synthesis community \cite{shirali2023better, pandey25_ssw, shirali2026naturalness}. Also, previous studies have shown that subjective naturalness ratings by humans can be influenced by various factors, including the wording of evaluation instructions \cite{dall14_speechprosody, kirkland23_ssw} and the characteristics of the evaluators
\cite{janska10_interspeech, dong11_blizzard, huang2026codecmos, ren2026mos}. 
Related issues have also been discussed in the perception of non-native speech, where closely related evaluation criteria such as accentedness, comprehensibility, and intelligibility can lead to different judgments of the same speech \cite{munro1995foreign, Derwing_Munro_1997}. In addition, intelligibility has been shown to vary depending on the listener's native language \cite{bent2003interlanguage}.
These findings suggest that the relationship between foreign accent and perceived naturalness may depend on how naturalness is defined and who performs the evaluation.

Motivated by these observations, this study investigates the effect of foreign accents on human naturalness evaluations, along with a comparison with automatic evaluation systems.
Specifically, we address the following research questions:
\begin{itemize}[leftmargin=*, nosep]
\item \textbf{RQ1:} Is the tendency to assign lower naturalness scores to speech with stronger foreign accents also observed in human subjective evaluations?
\item \textbf{RQ2:} How do evaluation instructions and evaluator backgrounds influence the relationship between foreign accent and perceived naturalness?
\end{itemize}
To answer these questions, we conducted a series of human and automatic evaluation experiments using synthetic speech samples with varying degrees of foreign accent\footnote{In this work, to enable a more controlled analysis of accent strength, we used analysis-resynthesized speech derived from real accented speech rather than speech generated by TTS systems.}. The main contribution of this paper is a large-scale subjective evaluation study of foreign-accented synthetic speech, involving 320 participants across multiple evaluation instructions and rater groups. 
The resulting findings are summarized as follows:
\begin{itemize}[leftmargin=*, nosep]
\item Human listeners also took accentedness into account when evaluating speech naturalness.
\item Human-likeness emerged as a more suitable evaluation criterion for foreign-accented synthetic speech, as it reduced the influence of accent while preserving sensitivity to system-level quality differences.
\item Raters with greater exposure to the target accent appeared to be better able to distinguish accent-related characteristics from system-induced unnaturalness.
\end{itemize}
Based on these findings, this study aims to open up discussion toward fairer and more appropriate naturalness evaluation of synthetic speech with foreign accents.

\section{Definition of Naturalness}
\label{sec:definition}
The ambiguity of the term \textit{naturalness} has often been criticized even in conventional speech synthesis evaluation, where foreign-accented speech is not specifically focused on\cite{cooper2024review}. Although naturalness-based MOS evaluation has served as an effective assessment method despite this ambiguity, recent studies have begun to reconsider what \textit{naturalness} actually refers to by breaking it down into finer components.

In \cite{pandey25_ssw}, the authors argued that \textit{naturalness} can be broadly divided into \textit{human-likeness} and \textit{appropriateness}. In conventional TTS research, \textit{naturalness} has been implicitly interpreted as being almost synonymous with \textit{human-likeness}. However, as advances in technology have enabled applications in a wider variety of contexts, evaluating whether a system meets its intended requirements has become difficult based solely on how human-like it sounds. As a result, researchers have begun to evaluate \textit{appropriateness} with respect to each application, such as fictional characters, robots, and situations where inclusivity regarding gender or accent is required.

A similar categorization is also discussed in \cite{nussbaum2025understanding}: \textit{human-likeness-based naturalness} and \textit{deviation-based naturalness}. 
Deviation-based naturalness measures naturalness as the degree of deviation from a reference representing maximum naturalness, with both the reference and the type of deviation explicitly specified. The reference may be either specific speech samples or an internal representation held by the rater.

Here, both \textit{appropriateness} and \textit{deviation-based naturalness} can be regarded as evaluations of conformity to expectations defined by a particular context. In the case of \textit{appropriateness}, such expectations are derived from application-specific requirements or social norms, whereas in \textit{deviation-based naturalness}, they are derived from a designated reference. From this perspective, we bring these concepts together under the notion of \textit{context conformity}, namely the degree to which speech conforms to context-dependent expectations. Together with \textit{human-likeness}, this leads to two perspectives on naturalness: 1) \textit{human-likeness} and 2) \textit{context conformity}.
We investigate how subjective evaluations of synthetic foreign-accented speech are affected when naturalness is more explicitly defined based on this categorization.

\section{Data Collection: How Do Humans Evaluate Synthetic Speech with a Foreign Accent?}
\label{sec:data}
To investigate how humans perceive and evaluate the \textit{naturalness} of foreign-accented synthetic speech, we conducted a subjective evaluation experiment. Speech samples for evaluation were generated by analysis-resynthesis of both native and non-native speech from the ERJ corpus \cite{Minematsu2004DevelopmentOE}. In this study, human raters were asked to evaluate each sample under a variety of conditions. In total, 320 participants were recruited, and 15,360 annotations were collected for 480 speech samples.

\subsection{Speech Stimuli}
\label{subsec:stimuli}
\noindent
\textbf{Selection of original speech samples:}
Samples from the phoneme-balanced sentence set (\texttt{S\_PH\_B\_1}) of the ERJ corpus were used as source speech for analysis-resynthesis.
Five target sentences
were selected from the set with the aim of maximizing phoneme coverage based on the CMU Pronouncing Dictionary\cite{weide1993cmudict}. Speech recordings from 24 speakers (20 Japanese learners of English and 4 native speakers of General American) were used, totaling 120 samples.

\noindent
\textbf{Speech restoration:}
The Japanese English (JE) speech in the ERJ corpus was collected at universities across Japan, resulting in variability in recording conditions. In contrast, the American English (AE) speech was recorded under studio conditions. Consequently, the JE recordings tend to be noisier than the AE recordings. To minimize the influence of the recording conditions, speech restoration was applied to all recordings before speech resynthesis. For this purpose, we used Sidon \cite{nakata2026sidon}. TABLE \ref{tab:restoration} presents mean DNSMOS \cite{dnsmos} scores for JE and AE, as well as the Spearman's rank correlation between the speaker-level mean UTMOS scores and the proficiency scores, before and after restoration. 
Before restoration, JE had lower DNSMOS than AE, whereas the scores became nearly identical after restoration.
Also, the correlation between UTMOS and proficiency
became stronger after restoration. The restoration process is thought to have mitigated the influence of noise, allowing accent strength to be more directly reflected in the predicted MOS scores.

\noindent
\textbf{Stimulus construction with analysis-resynthesis:}
%
The restored speech samples were then processed to construct four versions of each of the 120 original recordings: 
(1) a restoration-only version (No resyn.), and three versions further resynthesized with (2) BigVGAN \footnote{\url{https://hf.co/nvidia/bigvgan_v2_44khz_128band_512x}} \cite{lee2023bigvgan}, (3) EnCodec\footnote{\url{https://hf.co/facebook/encodec_24khz}, bandwidth set to 12.0 kbps.} \cite{fossez2023high}, and (4) WORLD \cite{morise2016world}. 
This resulted in an evaluation set of 480 speech stimuli. Since the output sampling rates varied across the analysis-resynthesis systems, all speech samples were resampled to 16,000 Hz prior to evaluation.

\begin{table}[tb]
	\centering
	\caption{Effects of speech restoration on ERJ corpus recordings: DNSMOS scores and Spearman's rank correlation coefficients between UTMOS and proficiency scores.}
	\label{tab:restoration}
    \vspace*{-2mm}
    \resizebox{0.75\columnwidth}{!}{
    \begin{tabular}{l|cc|c}
	  \toprule
		& \multicolumn{2}{c|}{DNSMOS} & UTMOS-Prof.   \\
          & JE & AE                &  correlation     \\
		\midrule   
       Original & 3.16 & 3.26 & 0.417 \\
       Restored & \textbf{3.30} & \textbf{3.34} & \textbf{0.549} \\
	  \bottomrule
	\end{tabular}
    }
    \vspace*{-5mm}
\end{table}

\subsection{Subjective evaluation design}
\label{subsec:disign}

\noindent
\textbf{Evaluation instructions:}
In this experiment, we investigate how subjective evaluations of synthetic speech with a foreign accent are affected when the intended meaning of \textit{naturalness} is explicitly specified in different ways. Based on the two categories of naturalness discussed in Sec. \ref{sec:definition}, 1) human-likeness and 2) context conformity, we prepared the four evaluation instructions shown in TABLE \ref{tab:instruction}. 
The first instruction (I1) is the simplest, yet also the most ambiguous. It asks listeners to rate how natural the speech sounds without providing any further definition of \textit{naturalness}. The second instruction (I2) is based on the definition of naturalness in terms of human-likeness (HL) and asks listeners to rate how human-like the speech sounds. The third and fourth instructions (I3, I4) are based on context conformity (CC). Listeners are asked to evaluate the naturalness of the speech under two different contextual assumptions: that the speech is produced by a native speaker or by a non-native speaker. 
For each instruction, listeners rated the speech samples on a five-point scale. To promote a consistent interpretation of the rating scale, each numerical score was accompanied by a descriptive label following \cite{kirkland23_ssw}: 1 (Bad), 2 (Poor), 3 (Fair), 4 (Good), and 5 (Excellent).
\begin{table}[tb]
	\centering
	\caption{Four evaluation instructions used in the subjective listening test. HL: Human-Likeness; CC: Context Conformity.}
	\label{tab:instruction}
    \vspace*{-2mm}
    \resizebox{\columnwidth}{!}{
    \begin{tabular}{c|c|l}
	  \toprule
		   & Type & ``Listen to the speech sample and rate ..." \\
        \midrule
       I1 & - & how natural it sounds.  \\
       I2 &  HL & how human-like (natural as human speech) it sounds.  \\
       I3 & CC & how natural it sounds as native speech.  \\
       I4 &  CC & how natural it sounds as non-native speech.  \\
	  \bottomrule
	\end{tabular}
    }
    \vspace*{-6mm}
\end{table}

\begin{table*}[tb]
	\centering
	\caption{Summary statistics of the constructed dataset: mean $\pm$ 95\% confidence interval (standard deviation). For each rater group, system, and speech type (AE or JE), MOS values lower and higher than those of I1 are shown in \textcolor{red}{red} and \textcolor{green!50!black}{green}, respectively. For each instruction, system, and speech type, the higher MOS values between native English and Japanese raters is shown in \textbf{bold}.}
	\label{tab:summary}
     \vspace*{-2mm}
    \resizebox{\textwidth}{!}{
    \begin{tabular}{ll|cc|cc|cc|cc}
	  \toprule
		& &\multicolumn{2}{c|}{No resyn.} & \multicolumn{2}{c|}{BigVGAN} &\multicolumn{2}{c|}{Encodec} & \multicolumn{2}{c}{WORLD} \\
     Rater   & & AE (N=20)& JE (N=100)& AE (N=20)& JE (N=100)& AE (N=20)& JE (N=100)& AE(N=20) & JE (N=100)\\
		\midrule   
        \multirow{4}{*}{EN} & I1 & \textbf{4.28$\pm$0.22} (0.49) & 2.75$\pm$0.15 (0.76) & 4.05$\pm$0.25 (0.56)& 2.71$\pm$0.14 (0.70)& \textbf{4.10$\pm$0.23} (0.52) & 2.58$\pm$0.14 (0.69)& \textbf{3.69$\pm$0.31} (0.71)& 2.44$\pm$0.13 (0.69)\\
        & I2 & \cellcolor{red!12}3.95$\pm$0.24 (0.55)&\cellcolor{green!12}2.89$\pm$0.14 (0.71) & \cellcolor{red!12}\textbf{4.01$\pm$0.26} (0.60)& \cellcolor{green!12}2.75$\pm$0.13 (0.68) & \cellcolor{red!12}3.80$\pm$0.23 (0.52)&\cellcolor{green!12}2.61$\pm$0.14 (0.70) & \cellcolor{red!12}3.29$\pm$0.27 (0.61)& \cellcolor{red!12}2.32$\pm$0.11 (0.55) \\
        & I3 & \cellcolor{red!12}3.96$\pm$0.24 (0.54)& \cellcolor{red!12}2.51$\pm$0.13 (0.65)& \cellcolor{green!12}4.06$\pm$0.29 (0.66)&\cellcolor{red!12}2.42$\pm$0.12 (0.62) &\cellcolor{red!12}3.93$\pm$0.23 (0.53) &\cellcolor{red!12}2.32$\pm$0.12 (0.59) &\cellcolor{red!12}3.53$\pm$0.26 (0.59) & \cellcolor{red!12}2.22$\pm$0.12 (0.63)\\
        & I4 & \cellcolor{red!12}\textbf{3.96$\pm$0.26} (0.59)&\cellcolor{green!12}2.91$\pm$0.12 (0.63) & \cellcolor{red!12}\textbf{3.83$\pm$0.25} (0.57)& \cellcolor{green!12}2.82$\pm$0.12 (0.63)& \cellcolor{red!12}\textbf{3.79$\pm$0.22} (0.50)&\cellcolor{green!12}2.70$\pm$0.12 (0.61) & \cellcolor{red!12}\textbf{3.34$\pm$0.27} (0.62)&\cellcolor{green!12}2.50$\pm$0.12 (0.61) \\
    \midrule   
        \multirow{4}{*}{JP} & I1 & 4.23$\pm$0.26 (0.59)& \textbf{3.33$\pm$0.12} (0.62) & \textbf{4.16$\pm$0.21} (0.48)& \textbf{3.24$\pm$0.12} (0.62)& 4.06$\pm$0.23 (0.52)& \textbf{3.14$\pm$0.13} (0.66)& 3.49$\pm$0.23 (0.52) &\textbf{2.93$\pm$0.13} (0.64)\\
        & I2 &\cellcolor{red!12}\textbf{4.00$\pm$0.27} (0.62)&\cellcolor{green!12}\textbf{3.65$\pm$0.11} (0.54) & \cellcolor{red!12}3.79$\pm$0.26 (0.60)&\cellcolor{green!12}\textbf{3.39$\pm$0.11} (0.57) & \cellcolor{red!12}\textbf{3.81$\pm$0.25} (0.58)&\cellcolor{green!12}\textbf{3.27$\pm$0.12} (0.61) & \cellcolor{red!12}\textbf{3.39$\pm$0.32} (0.74)&\cellcolor{green!12}\textbf{3.06$\pm$0.12} (0.60) \\
        & I3 & \cellcolor{red!12}\textbf{4.18$\pm$0.19} (0.44)& \cellcolor{red!12}\textbf{2.78$\pm$0.14} (0.70)& \cellcolor{red!12}\textbf{4.11$\pm$0.21} (0.47)& \cellcolor{red!12}\textbf{2.78$\pm$0.13} (0.65)& \cellcolor{red!12}\textbf{4.04$\pm$0.23} (0.53)&\cellcolor{red!12}\textbf{2.68$\pm$0.13} (0.68)& \cellcolor{green!12}\textbf{3.91$\pm$0.24} (0.55)&\cellcolor{red!12}\textbf{2.70$\pm$0.13} (0.67) \\
        & I4 & \cellcolor{red!12}3.54$\pm$0.30 (0.69)& \cellcolor{red!12}\textbf{3.10$\pm$0.11} (0.55) & \cellcolor{red!12}3.26$\pm$0.30 (0.67)&\cellcolor{red!12}\textbf{3.05$\pm$0.11} (0.57)& \cellcolor{red!12}3.25$\pm$0.32 (0.72)&\cellcolor{red!12}\textbf{3.07$\pm$0.11} (0.55) &\cellcolor{red!12}3.29$\pm$0.31 (0.70) & \cellcolor{green!12}\textbf{2.97$\pm$0.11} (0.54)\\

	  \bottomrule
	\end{tabular}
    }
    \vspace*{-4mm}
\end{table*}

\noindent
\textbf{Stimulus grouping:}
To reduce the listening burden on individual raters, the 480 speech samples in the evaluation set described in Sec. \ref{subsec:stimuli} were divided into 10 subgroups of 48 samples each. Each subgroup contained all four resynthesis versions of 12 original utterances (10 JEs and 2 AEs). The 12 utterances within a subgroup were produced by different speakers, and the five target sentences were distributed across subgroups as evenly as possible. To ensure that the JE utterances in each subgroup covered a similar and wide range of proficiency levels, the 20 Japanese speakers were first sorted by proficiency score and then divided into odd- and even-ranked groups. Each subgroup was constructed using speakers from one of these two groups.

\noindent
\textbf{Crowdsourced listening test:}
Using the four evaluation instructions and the 10 subgroups described above, subjective evaluations were conducted via crowdsourcing platforms\footnote{We used Prolific (\url{https://www.prolific.com/}) for recruiting native English speakers and Lancers (\url{https://www.lancers.jp}) for native Japanese speakers.}. For each instruction-subgroup combination, four native speakers of English and four native speakers of Japanese (i.e., listeners who shared the same native language as the speakers of the non-native evaluation stimuli) were recruited\footnote{All participants provided informed consent before the listening test.}.  
For quality control, two highly unnatural speech samples were inserted into each evaluation set. These samples were generated using WORLD by fixing the f0 to a constant value throughout the utterance.
Only listeners who assigned low ratings (i.e., 1 or 2) to both samples were retained for analysis, and recruitment continued until the required number of valid responses was obtained for each instruction-subgroup combination.
Consequently, 15,360 ratings provided by 320 listeners (160 native speakers of English and 160 native speakers of Japanese) were included in the final analysis dataset\footnote{Resources for reproducing the listening-test stimuli and the evaluation results are available at \url{https://github.com/ondatk68/foreign-accent-mos/}}.

\section{Data Analysis}
\subsection{Overall results}
\label{subsec:overall}
TABLE \ref{tab:summary} summarizes the results of the subjective evaluation described in Sec. \ref{sec:data}. These results indicate several consistent patterns in human evaluations of the \textit{naturalness} of synthetic speech with foreign accents.

\noindent \textbf{Effect of accents:} Across all systems, evaluation instructions, and rater groups, AE consistently received higher ratings than JE. This tendency was observed even under I4, which explicitly asked listeners to evaluate ``naturalness as non-native speech". This result suggests that human listeners strongly associate foreign accents with reduced naturalness, regardless of the evaluation criteria provided. In addition, JE showed higher standard deviations than AE in most cases, suggesting that variability in accentedness was reflected in MOS values.

\noindent \textbf{Effect of evaluation instructions:} The absolute MOS values were found to vary depending on the evaluation instruction. Compared with I1, which simply asked listeners to evaluate ``naturalness", under I2, where they are explicitly directed to assess ``human-likeness", listeners generally assigned lower ratings to AE and higher ratings to JE across both rater groups. As a result, the rating gap between AE and JE was reduced, suggesting a reduced influence of foreign accent on perceived naturalness. Furthermore, under I3, which asked listeners to evaluate ``naturalness as native speech", JE received lower ratings than I1 in all cases. Conversely, under I4, which asked listeners to evaluate ``naturalness as non-native speech", AE was rated lower than I1. These results suggest that listeners adapt their evaluation criteria according to the instructions provided, particularly in how they account for foreign accent in naturalness judgments.

\noindent \textbf{Effect of raters' linguistic backgrounds:} Across all instruction conditions and systems, Japanese native raters consistently assigned higher ratings to JE than English native raters. This may be attributed to Japanese raters' greater exposure to Japanese-accented English, which could lead them to perceive such speech as more natural. In contrast, ratings for AE speech were generally similar across the two listener groups under I1 and I2. 
Although previous studies have shown that native raters are more sensitive than non-native raters to unnatural prosody introduced by TTS systems\cite{janska10_interspeech, dong11_blizzard}, the present analysis-resynthesis setting largely preserved human-produced segmental and prosodic characteristics. This may explain why no substantial difference was observed between rater groups.

\noindent \textbf{System ranking:} Despite differences in listener backgrounds and evaluation instructions, the relative ranking of the four resynthesis systems remained largely consistent, for both AE and JE. No resyn. and BigVGAN generally achieved the highest ratings, followed by Encodec and WORLD. 
Nevertheless, some evaluation conditions produced substantially smaller rating gaps among systems, suggesting that the discriminability of system quality depends on how the evaluation task is framed. We discuss this observation in detail in Sec. \ref{subsec:system}.

\subsection{Relationship between accent strength and perceived naturalness}
\label{subsec:rel}

\begin{figure*}[t]
  \centering
  \includegraphics[width=0.90\linewidth]{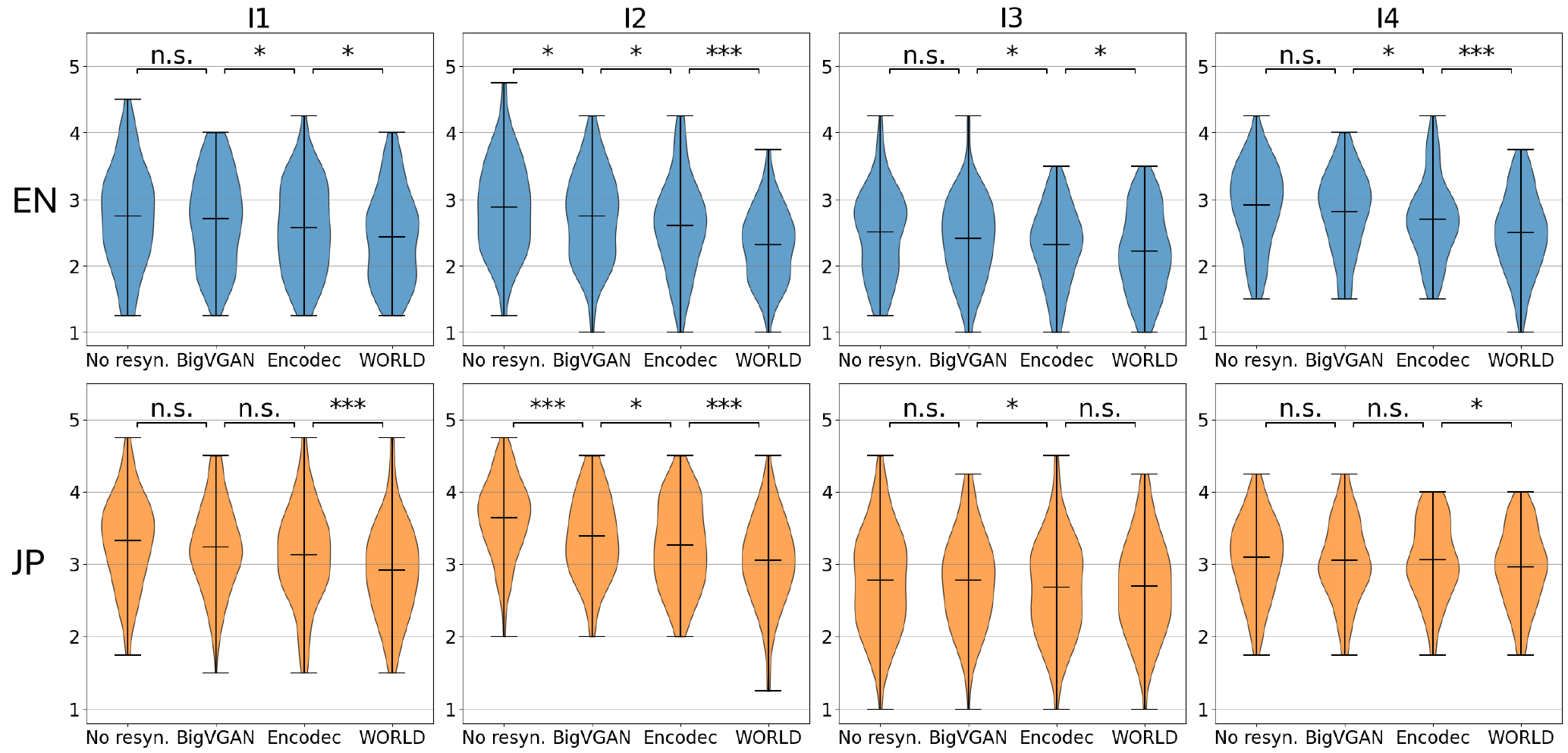}
  \vspace*{-3.5mm}
  \caption{Violin plots of MOS values for JE speech by rater group and evaluation instruction. Statistical significance was assessed using Wilcoxon signed-rank tests with Holm correction (*: $p<0.05$, **: $p<0.01$, ***: $p<0.001$, n.s.: not significant).}
  \label{fig:violin}
  \vspace*{-5.5mm}
\end{figure*}

To further examine how foreign accent affects human naturalness judgments, we analyzed the relationship between speakers' proficiency scores and MOS values within the JE speech samples. 
TABLE \ref{tab:spearman} shows Spearman's rank correlation coefficients between proficiency scores and speaker-level average of MOS values. 
We used each speaker’s ERJ proficiency score for all corresponding resynthesized samples.
Overall, positive correlations were observed across all conditions. This indicates that speech from speakers with weaker accents (= higher proficiency) tended to be judged as more natural, consistent with the UTMOS results in Fig.\ref{fig:erj-utmos}.
However, the strength of this tendency was found to vary depending on the evaluation instruction and the raters' linguistic backgrounds.

\noindent \textbf{Naturalness vs. human-likeness:} Comparing I1 (naturalness) and I2 (human-likeness), correlations were generally weaker under I2 in almost all cases. Together with the higher JE ratings observed under I2 in TABLE \ref{tab:summary}, this suggests that focusing listeners' attention on human-likeness reduces the influence of accentedness on naturalness evaluations. Although the two rater groups showed similar correlations under I1, native English raters showed higher correlations than native Japanese raters under I2. This may suggest that, because native English listeners have less exposure to Japanese-accented English, 
they are more likely to interpret phonetic and prosodic deviations from native speech as system-induced unnaturalness rather than as accent-related characteristics.
As a result, even under human-likeness criterion, they showed a stronger tendency to assign lower scores to speech with stronger accents.

\begin{table}[tb]
	\centering
	\caption{Spearman's rank correlation coefficients between MOS scores and proficiency scores.}
	\label{tab:spearman}
    \vspace*{-2mm}
    \resizebox{\columnwidth}{!}{
    \begin{tabular}{ll|cccc|c}
	  \toprule
		Rater& & No resyn. & BigVGAN & Encodec & WORLD & Avg. \\
        \midrule
        EN &I1 & 0.455& 0.535&0.549 & 0.521 & 0.515\\
        &I2 & 0.588&0.526 &0.392 &0.202 & 0.427 \\
        &I3 & 0.446& 0.317& 0.469& 0.426 & 0.415\\
        &I4 & 0.381&0.259 &0.308 &0.484 & 0.358 \\
        \midrule
        JP &I1 & 0.575& 0.437& 0.641& 0.411 & 0.516\\
        &I2 &0.491 &0.284 &0.270 &0.254 & 0.325 \\
        &I3 & 0.498& 0.461& 0.461& 0.417 & 0.459\\
        &I4 & 0.350& 0.239& 0.268& 0.237 & 0.274\\

	  \bottomrule
	\end{tabular}
    }
    \vspace*{-5mm}
\end{table}

\noindent \textbf{Effect of non-native-speech context:} Among the four instruction conditions, I4 (naturalness as non-native speech) generally showed the weakest correlations. This suggests that providing listeners with a non-native speech context made them less likely to associate stronger accents with lower naturalness. As in I2, the correlations were also lower for Japanese raters than for native English raters, possibly due to their greater familiarity with Japanese accents. However, positive correlations remained in all cases, indicating that although this evaluation framework can reduce the effect of speaker proficiency on perceived naturalness, it cannot eliminate it entirely.

\subsection{Can humans evaluate systems beyond foreign accent?}
\label{subsec:system}
So far, we have examined how the presence and degree of foreign accent affect naturalness evaluations. In this section, we discuss whether listeners can accurately evaluate system-level quality differences in the presence of foreign accents. Fig. \ref{fig:violin} shows violin plots of MOS values for JE speech. Following \cite{clark07_blizzard}, statistical significance between systems was assessed using Wilcoxon signed-rank tests. To account for multiple comparisons, Holm correction was applied.

As an overall trend, system-level differences were most clearly observed under I2 (human-likeness). As shown in Secs. \ref{subsec:overall} and \ref{subsec:rel}, explicitly directing listeners to assess human-likeness appears to reduce the influence of foreign accent. This, in turn, enables more reliable evaluation of system-level differences even for accented speech. 
In contrast, under I3 and I4, system-level differences became less apparent, particularly among Japanese raters. This may be because explicitly framing the evaluation within a native or non-native speech context directed listeners' attention toward how well the speech conformed to that context, making differences between synthesis systems less noticeable. Although I4 exhibited the weakest proficiency-MOS correlations in TABLE \ref{tab:spearman}, this does not necessarily imply that listeners focused more on system-level quality. Rather, listeners may have paid greater attention to accentedness itself, but a stronger perceived accent was not necessarily associated with lower naturalness. This reduced the proficiency-MOS correlation, while at the same time making differences between synthesis systems less apparent.

Based on the analyses presented so far, \textit{human-likeness} appears to be the most suitable evaluation criterion for foreign-accented synthetic speech. Compared with general naturalness, the human-likeness criterion reduced the influence of accentedness and showed the clearest system-level differences among the four evaluation instructions tested. Also, the results suggest that Japanese raters, who had greater exposure to Japanese-accented English, may have been better able to distinguish accent-related characteristics from system-induced unnaturalness when judging human-likeness.

\subsection{How do existing automatic evaluation systems evaluate foreign-accented synthetic speech?}
\begin{figure}[tb]
  \centering
  \includegraphics[width=\columnwidth]{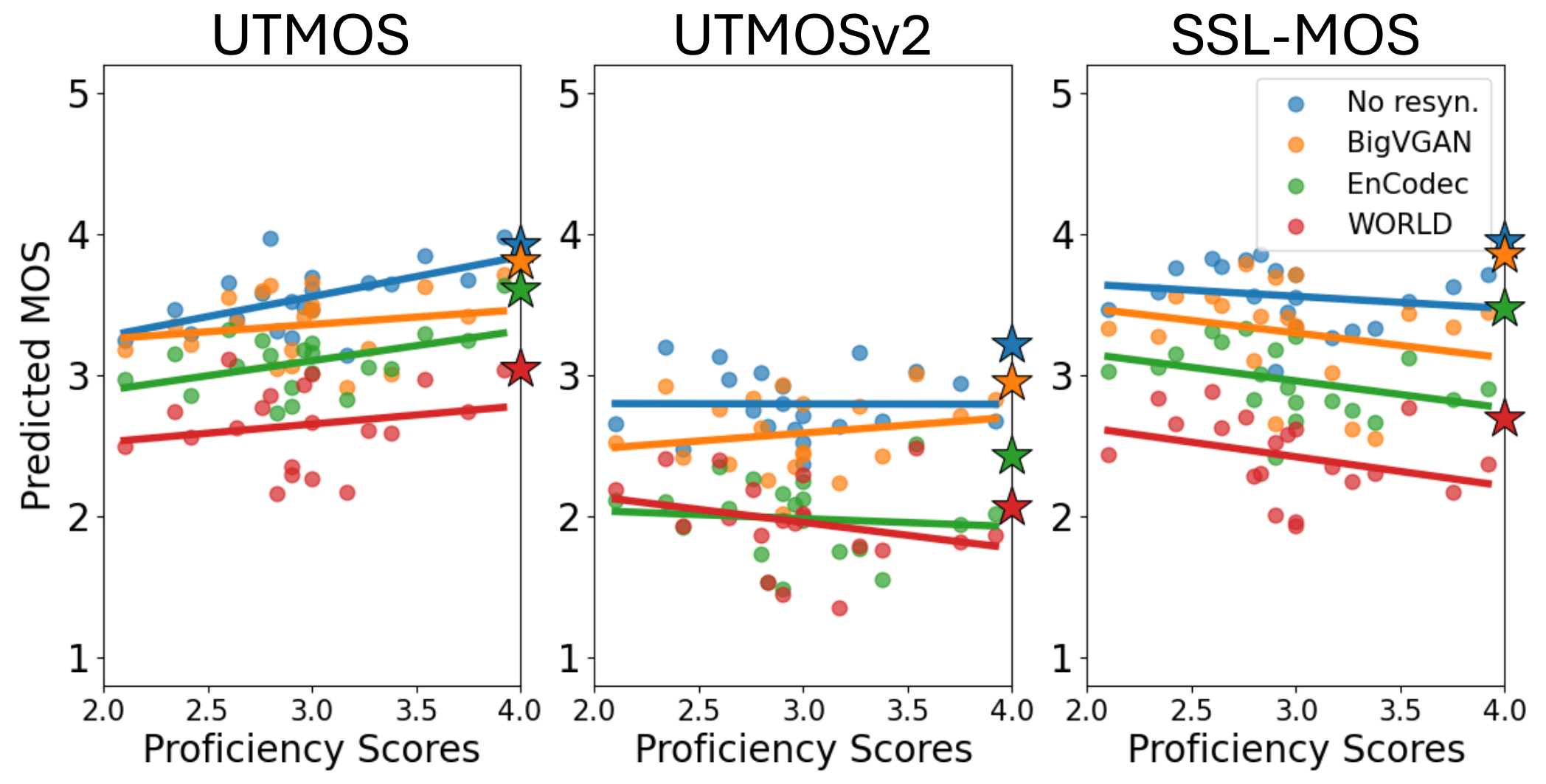}
  \vspace*{-8mm}
  \caption{Relationship between proficiency scores and speaker-level mean pseudo MOS scores predicted by each system for JE speech. Stars indicate the overall mean pseudo MOS scores for AE speech.}
  \label{fig:autmatic}
  \vspace*{-3mm}
\end{figure}

Lastly, to examine how existing MOS prediction models behave when evaluating foreign-accented synthetic speech, we evaluated the dataset constructed in this study using three models: UTMOS\cite{saeki22c_interspeech}, UTMOSv2\cite{baba2024t05}, and SSL-MOS\footnote{We used the default model provided by SHEET toolkit (v0.2.5)\cite{huang25g_interspeech}.}\cite{cooper2022sslmos}. Fig. \ref{fig:autmatic} shows scatter plots of the speaker-level means of the scores predicted by each model against the proficiency scores for JE speech. For reference, the mean predicted scores for AE speech are indicated with stars. All models tended to assign higher scores to AE than to JE, consistent with the human evaluations. However, the trends in relation to speaker proficiency varied across models. Consistent with Fig. \ref{fig:erj-utmos}, UTMOS exhibited a positive correlation with proficiency scores, whereas SSL-MOS somehow showed a negative correlation. Regarding system-level comparison, significant differences were found between all system pairs for UTMOS and SSL-MOS ($p<0.001$). In contrast, UTMOSv2 showed no significant difference between EnCodec and WORLD. 
TABLE \ref{tab:spearman-auto} presents the utterance-level Spearman's rank correlation coefficients between the scores predicted by each model and the human MOS values for JE speech. Interestingly, all three models showed the highest correlations with the ratings obtained under I2 (human-likeness). This suggests that existing MOS prediction models may, to some extent, evaluate speech in a manner consistent with human-likeness judgments. Nevertheless, this does not imply that existing evaluation metrics can be directly applied to foreign-accented synthetic speech as is. The sensitivity to accentedness observed in both human and automatic evaluations suggests that accent-related bias remains a fundamental challenge in assessing the naturalness of foreign-accented synthetic speech.

\begin{table}[tb]
	\centering
	\caption{Utterance-level Spearman's rank correlation coefficients between the scores predicted by each MOS prediction system and the actual MOS values (JE only, N=400).}
	\label{tab:spearman-auto}
    \vspace*{-2mm}
    \resizebox{\columnwidth}{!}{
    \begin{tabular}{l|cccc|cccc}
	  \toprule
		& \multicolumn{4}{c|}{EN}& \multicolumn{4}{c}{JP}\\
        & I1 & I2 & I3 & I4 & I1 & I2 & I3 & I4 \\
        \midrule
        UTMOS & 0.26 & \textbf{0.42} & 0.28 & 0.33 & 0.49 & \textbf{0.51} & 0.28 & 0.24 \\
        UTMOSv2 & 0.15 & \textbf{0.33} & 0.26 & 0.17 & 0.28 & \textbf{0.34} & 0.23 & 0.07 \\
        SSL-MOS & 0.14 & \textbf{0.28} & 0.17 & 0.19 & 0.27 & \textbf{0.34} & -0.01 & 0.03 \\ 
	  \bottomrule
	\end{tabular}
    }
    \vspace*{-4mm}
\end{table}

\section{Discussion: Implications for \\Evaluating Accented Speech Synthesis}
The results of this study indicate that the evaluation of foreign-accented synthetic speech involves a fundamental difficulty that is not limited to automatic evaluation systems. Across all conditions, JE speech received lower ratings than AE speech, and within JE speech, proficiency scores were positively correlated with MOS values. These results suggest that human listeners also associate accentedness with reduced naturalness.
The human-likeness instruction mitigated the influence of accentedness while making system-level differences more clearly observable. However, the effect was not complete. 
Even under the human-likeness criterion, native English raters showed a stronger tendency to assign lower scores to speech with stronger accents, compared to Japanese raters.
One possible explanation is that listeners who are less familiar with the target accent may be more likely to interpret accent-related phonetic and prosodic characteristics as system-induced unnaturalness. This suggests that appropriate evaluation of accented synthetic speech requires both carefully designed instructions and rater selection based on familiarity with the target accent.

A further point to note is that the present study was conducted under a highly controlled analysis-resynthesis setting. The degree of accent was largely controlled across systems, and the segmental and prosodic characteristics of the stimuli were mostly inherited from real human speech. 
However, the evaluation of accented TTS is likely to be even more challenging. In TTS, the degree of accent may differ across systems, and the synthesis system may introduce segmental or prosodic unnaturalness independently of the intended accent characteristics. Under such conditions, it becomes even more difficult to determine whether a perceived deviation should be attributed to the intended foreign accent or to a system defect. 

Therefore, both instruction design and rater selection will be increasingly important in future evaluations of accented speech synthesis. Raters should have sufficient exposure to the target accent and be able to appropriately distinguish accent-related characteristics from system-induced unnaturalness. 
In this sense, the results may also point to a limitation of crowdsourced subjective evaluation. Speech synthesis systems can now generate diverse speech types beyond conventional neutral native speech, for which many listeners share relatively common naturalness criteria. As synthesis targets become more diverse, evaluation protocols also need to become more context-specific and may require judgments from listeners with expertise in the target speech characteristics.

\section{Conclusions}
In this study, we conducted a large-scale subjective evaluation to investigate how human listeners evaluate foreign-accented synthetic speech. Based on two categories of naturalness, human-likeness and context conformity, we designed four evaluation instructions. A total of 320 raters, including both native English and native Japanese speakers, took part in the experiment, providing 15,360 five-point ratings in total.

The analysis of the results highlighted the importance of both evaluation instructions and rater selection for appropriately evaluating accented synthetic speech. It also suggested limitations of conventional evaluations based simply on “naturalness,” as well as those relying solely on crowdsourced listening tests.
We hope that the analyses and discussions presented in this paper provide useful insights and serve as a starting point for further discussion on how foreign-accented synthetic speech should be evaluated.

\section*{Acknowledgment}
This paper is based on results obtained from the BRIDGE Program (R7-H05), implemented by the Cabinet Office, Government of Japan.
This work was supported by JST SPRING, Grant Number JPMJSP2108.

\section*{AI-Generated Content Disclosure}
We used ChatGPT (OpenAI) to refine English expressions throughout the manuscript.

\bibliographystyle{IEEEtran}
\bibliography{references}

@article{flege1995second,
	author = {Flege, James E},
	journal = {Speech perception and linguistic experience: Issues in cross-language research},
	pages = {233--277},
	publisher = {York Press},
	title = {Second language speech learning: Theory, findings, and problems},
	volume = {92},
	year = {1995}}

@inproceedings{onda24_interspeech,
  title     = {A Pilot Study of {GSLM}-based Simulation of Foreign Accentuation Only Using Native Speech Corpora},
  author    = {Kentaro Onda and Joonyong Park and Nobuaki Minematsu and Daisuke Saito},
  year      = {2024},
  booktitle = {Interspeech 2024},
  pages     = {3600--3604},
  doi       = {10.21437/Interspeech.2024-1473},
  issn      = {2958-1796},
}

@inproceedings{Minematsu2004DevelopmentOE,
  title={Development of {E}nglish Speech Database Read by {J}apanese to Support {CALL} Research},
  author={Nobuaki Minematsu and Yoshihiro Tomiyama and Kei Yoshimoto and Katsumasa Shimizu and Seiichi Nakagawa and Masatake Dantsuji and Shozo Makino},
  year={2004},
  booktitle={ICA 2004},
  pages={557-560},
  _url={https://api.semanticscholar.org/CorpusID:6642302}
}

@article{munro1995foreign,
	title={Foreign accent, comprehensibility, and intelligibility in the speech of second language learners},
	author={Munro, Murray J and Derwing, Tracey M},
	journal={Language Learning},
	volume={45},
	number={1},
	pages={73--97},
	year={1995},
	publisher={Wiley Online Library}
}

@article{bent2003interlanguage,
  title={The interlanguage speech intelligibility benefit},
  author={Bent, Tessa and Bradlow, Ann R},
  _journal={The Journal of the Acoustical Society of America},
  journal={JASA},
  volume={114},
  number={3},
  pages={1600--1610},
  year={2003},
  publisher={Acoustical Society of America}
}

@inproceedings{chang24b_interspeech,
  title     = {The {I}nterspeech 2024 Challenge on Speech Processing Using Discrete Units},
  author    = {Xuankai Chang and Jiatong Shi and Jinchuan Tian and Yuning Wu and Yuxun Tang and Yihan Wu and Shinji Watanabe and Yossi Adi and Xie Chen and Qin Jin},
  year      = {2024},
  booktitle = {Interspeech 2024},
  pages     = {2559--2563},
  doi       = {10.21437/Interspeech.2024-1878},
  issn      = {2958-1796},
}

@article{morise2016world,
  title={WORLD: a vocoder-based high-quality speech synthesis system for real-time applications},
  author={Morise, Masanori and Yokomori, Fumiya and Ozawa, Kenji},
  journal={IEICE TRANSACTIONS on Information and Systems},
  volume={99},
  number={7},
  pages={1877--1884},
  year={2016},
  publisher={The Institute of Electronics, Information and Communication Engineers}
}

@article{fossez2023high,
title={High Fidelity Neural Audio Compression},
author={Alexandre D{\'e}fossez and Jade Copet and Gabriel Synnaeve and Yossi Adi},
journal={Transactions on Machine Learning Research},
issn={2835-8856},
year={2023},
_url={https://openreview.net/forum?id=ivCd8z8zR2},
_note={Featured Certification, Reproducibility Certification}
}

@inproceedings{saeki22c_interspeech,
  title     = {{UTMOS: UTokyo-SaruLab System for VoiceMOS Challenge 2022}},
  author    = {Takaaki Saeki and Detai Xin and Wataru Nakata and Tomoki Koriyama and Shinnosuke Takamichi and Hiroshi Saruwatari},
  year      = {{2022}},
  booktitle = {{Interspeech 2022}},
  pages     = {{4521--4525}},
  doi       = {{10.21437/Interspeech.2022-439}},
  issn      = {{2958-1796}},
}

@article{tan2021survey,
  title={A survey on neural speech synthesis},
  author={Tan, Xu and Qin, Tao and Soong, Frank and Liu, Tie-Yan},
  journal={arXiv preprint arXiv:2106.15561},
  year={2021}
}

@article{tan2024naturalspeech,
  title={Naturalspeech: End-to-end text-to-speech synthesis with human-level quality},
  author={Tan, Xu and Chen, Jiawei and Liu, Haohe and Cong, Jian and Zhang, Chen and Liu, Yanqing and Wang, Xi and Leng, Yichong and Yi, Yuanhao and He, Lei and others},
  journal={IEEE Transactions on Pattern Analysis and Machine Intelligence},
  volume={46},
  number={6},
  pages={4234--4245},
  year={2024},
  publisher={IEEE}
}

@article{chen2024vall,
  title={Vall-e 2: Neural codec language models are human parity zero-shot text to speech synthesizers},
  author={Chen, Sanyuan and Liu, Shujie and Zhou, Long and Liu, Yanqing and Tan, Xu and Li, Jinyu and Zhao, Sheng and Qian, Yao and Wei, Furu},
  journal={arXiv preprint arXiv:2406.05370},
  year={2024}
}

@inproceedings{xie-etal-2025-towards,
    title = "Towards Controllable Speech Synthesis in the Era of Large Language Models: A Systematic Survey",
    author = "Xie, Tianxin  and
      Rong, Yan  and
      Zhang, Pengfei  and
      Wang, Wenwu  and
      Liu, Li",
    _editor = "Christodoulopoulos, Christos  and
      Chakraborty, Tanmoy  and
      Rose, Carolyn  and
      Peng, Violet",
    booktitle = "Proceedings of the 2025 Conference on Empirical Methods in Natural Language Processing",
    _month = nov,
    year = "2025",
    _address = "Suzhou, China",
_publisher = "Association for Computational Linguistics",
    _url = "https://aclanthology.org/2025.emnlp-main.40/",
    doi = "10.18653/v1/2025.emnlp-main.40",
    pages = "764--791",
    ISBN = "979-8-89176-332-6",
    abstract = "Text-to-speech (TTS) has advanced from generating natural-sounding speech to enabling fine-grained control over attributes like emotion, timbre, and style. Driven by rising industrial demand and breakthroughs in deep learning, e.g., diffusion and large language models (LLMs), controllable TTS has become a rapidly growing research area. This survey provides **the first** comprehensive review of controllable TTS methods, from traditional control techniques to emerging approaches using natural language prompts. We categorize model architectures, control strategies, and feature representations, while also summarizing challenges, datasets, and evaluations in controllable TTS. This survey aims to guide researchers and practitioners by offering a clear taxonomy and highlighting future directions in this fast-evolving field. One can visit https://github.com/imxtx/awesome-controllabe-speech-synthesis for a comprehensive paper list and updates."
}

@inproceedings{onda25b_interspeech,
  title     = {{Prosodically Enhanced Foreign Accent Simulation by Discrete Token-based Resynthesis Only with Native Speech Corpora}},
  author    = {Kentaro Onda and Keisuke Imoto and Satoru Fukayama and Daisuke Saito and Nobuaki Minematsu},
  year      = {2025},
  booktitle = {{Interspeech 2025}},
  pages     = {2195--2199},
  doi       = {10.21437/Interspeech.2025-590},
  issn      = {2958-1796},
}

@inproceedings{inoue2025macst,
  title={Macst: Multi-accent speech synthesis via text transliteration for accent conversion},
  author={Inoue, Sho and Wang, Shuai and Wang, Wanxing and Zhu, Pengcheng and Bi, Mengxiao and Li, Haizhou},
  booktitle={ICASSP 2025-2025 IEEE International Conference on Acoustics, Speech and Signal Processing (ICASSP)},
  pages={1--5},
  year={2025},
  _organization={IEEE}
}

@article{lertpetchpun2026accent,
  title={Accent Vector: Controllable Accent Manipulation for Multilingual TTS Without Accented Data},
  author={Lertpetchpun, Thanathai and Trachu, Thanapat and Lee, Jihwan and Feng, Tiantian and Byrd, Dani and Narayanan, Shrikanth},
  journal={arXiv preprint arXiv:2603.07534},
  year={2026}
}

@inproceedings{yang2025emovoice,
  title={Emovoice: Llm-based emotional text-to-speech model with freestyle text prompting},
  author={Yang, Guanrou and Yang, Chen and Chen, Qian and Ma, Ziyang and Chen, Wenxi and Wang, Wen and Wang, Tianrui and Yang, Yifan and Niu, Zhikang and Liu, Wenrui and others},
  booktitle={Proceedings of the 33rd ACM International Conference on Multimedia},
  pages={10748--10757},
  year={2025}
}

@inproceedings{diatlova23_ssw,
  title     = {{EmoSpeech: guiding FastSpeech2 towards Emotional Text to Speech}},
  author    = {Daria Diatlova and Vitalii Shutov},
  year      = {2023},
  booktitle = {{12th ISCA Speech Synthesis Workshop (SSW2023)}},
  pages     = {106--112},
  doi       = {10.21437/SSW.2023-17},
}

@inproceedings{guo2023emodiff,
  title={Emodiff: Intensity controllable emotional text-to-speech with soft-label guidance},
  author={Guo, Yiwei and Du, Chenpeng and Chen, Xie and Yu, Kai},
  booktitle={ICASSP 2023-2023 IEEE International conference on acoustics, speech and signal processing (ICASSP)},
  pages={1--5},
  year={2023},
  _organization={IEEE}
}

@inproceedings{guo2023prompttts,
  title={Prompttts: Controllable text-to-speech with text descriptions},
  author={Guo, Zhifang and Leng, Yichong and Wu, Yihan and Zhao, Sheng and Tan, Xu},
  booktitle={ICASSP 2023-2023 IEEE International Conference on Acoustics, Speech and Signal Processing (ICASSP)},
  pages={1--5},
  year={2023},
  _organization={IEEE}
}

@article{yang2024instructtts,
  title={Instructtts: Modelling expressive tts in discrete latent space with natural language style prompt},
  author={Yang, Dongchao and Liu, Songxiang and Huang, Rongjie and Weng, Chao and Meng, Helen},
  journal={IEEE/ACM Transactions on Audio, Speech, and Language Processing},
  volume={32},
  pages={2913--2925},
  year={2024},
  publisher={IEEE}
}

@inproceedings{ji-etal-2025-controlspeech,
    title = "{C}ontrol{S}peech: Towards Simultaneous and Independent Zero-shot Speaker Cloning and Zero-shot Language Style Control",
    author = "Ji, Shengpeng  and
      Chen, Qian  and
      Wang, Wen  and
      Zuo, Jialong  and
      Fang, Minghui  and
      Jiang, Ziyue  and
      Huang, Hai  and
      Wang, Zehan  and
      Cheng, Xize  and
      Zheng, Siqi  and
      Zhao, Zhou",
    _editor = "Che, Wanxiang  and
      Nabende, Joyce  and
      Shutova, Ekaterina  and
      Pilehvar, Mohammad Taher",
    booktitle = "Proceedings of the 63rd Annual Meeting of the Association for Computational Linguistics (Volume 1: Long Papers)",
    _month = jul,
    year = "2025",
    _address = "Vienna, Austria",
    _publisher = "Association for Computational Linguistics",
    _url = "https://aclanthology.org/2025.acl-long.346/",
    doi = "10.18653/v1/2025.acl-long.346",
    pages = "6966--6981",
    ISBN = "979-8-89176-251-0",
    abstract = "In this paper, we present ControlSpeech, a text-to-speech (TTS) system capable of fully cloning the speaker{'}s voice and enabling arbitrary control and adjustment of speaking style. Prior zero-shot TTS models only mimic the speaker{'}s voice without further control and adjustment capabilities while prior controllable TTS models cannot perform speaker-specific voice generation. Therefore, ControlSpeech focuses on a more challenging task{---}a TTS system with controllable timbre, content, and style at the same time. ControlSpeech takes speech prompts, content prompts, and style prompts as inputs and utilizes bidirectional attention and mask-based parallel decoding to capture codec representations corresponding to timbre, content, and style in a discrete decoupling codec space. Moreover, we analyze the many-to-many issue in textual style control and propose the Style Mixture Semantic Density (SMSD) module, which is based on Gaussian mixture density networks, to resolve this problem. To facilitate empirical validations, we make available a new style controllable dataset called VccmDataset. Our experimental results demonstrate that ControlSpeech exhibits comparable or state-of-the-art (SOTA) performance in terms of controllability, timbre similarity, audio quality, robustness, and generalizability. Codes are available at https://github.com/jishengpeng/ControlSpeech."
}

@inproceedings{huang22f_interspeech,
  title     = {{The VoiceMOS Challenge 2022}},
  author    = {Wen Chin Huang and Erica Cooper and Yu Tsao and Hsin-Min Wang and Tomoki Toda and Junichi Yamagishi},
  year      = {2022},
  booktitle = {{Interspeech 2022}},
  pages     = {4536--4540},
  doi       = {10.21437/Interspeech.2022-970},
  issn      = {2958-1796},
}

@inproceedings{cooper2023voicemos,
  title={The VoiceMOS Challenge 2023: Zero-shot subjective speech quality prediction for multiple domains},
  author={Cooper, Erica and Huang, Wen-Chin and Tsao, Yu and Wang, Hsin-Min and Toda, Tomoki and Yamagishi, Junichi},
  booktitle={2023 IEEE Automatic Speech Recognition and Understanding Workshop (ASRU)},
  pages={1--7},
  year={2023},
  _organization={IEEE}
}

@inproceedings{huang2024voicemos,
  title={The VoiceMOS challenge 2024: Beyond speech quality prediction},
  author={Huang, Wen-Chin and Fu, Szu-Wei and Cooper, Erica and Zezario, Ryandhimas E and Toda, Tomoki and Wang, Hsin-Min and Yamagishi, Junichi and Tsao, Yu},
  booktitle={2024 IEEE Spoken Language Technology Workshop (SLT)},
  pages={803--810},
  year={2024},
  _organization={IEEE}
}

@inproceedings{saijo25_interspeech,
  title     = {{Interspeech 2025 URGENT Speech Enhancement Challenge}},
  author    = {Kohei Saijo and Wangyou Zhang and Samuele Cornell and Robin Scheibler and Chenda Li and Zhaoheng Ni and Anurag Kumar and Marvin Sach and Yihui Fu and Wei Wang and Tim Fingscheidt and Shinji Watanabe},
  year      = {2025},
  booktitle = {{Interspeech 2025}},
  pages     = {858--862},
  doi       = {10.21437/Interspeech.2025-1363},
  issn      = {2958-1796},
}

@article{li2026icassp,
  title={ICASSP 2026 URGENT Speech Enhancement Challenge},
  author={Li, Chenda and Wang, Wei and Sach, Marvin and Zhang, Wangyou and Saijo, Kohei and Cornell, Samuele and Fu, Yihui and Ni, Zhaoheng and Fingscheidt, Tim and Watanabe, Shinji and others},
  journal={arXiv preprint arXiv:2601.13531},
  year={2026}
}

@article{cooper2024review,
  title={A review on subjective and objective evaluation of synthetic speech},
  author={Cooper, Erica and Huang, Wen-Chin and Tsao, Yu and Wang, Hsin-Min and Toda, Tomoki and Yamagishi, Junichi},
  journal={Acoustical Science and Technology},
  volume={45},
  number={4},
  pages={161--183},
  year={2024},
  publisher={ACOUSTICAL SOCIETY OF JAPAN}
}

@inproceedings{tseng21b_interspeech,
  title     = {{Utilizing Self-Supervised Representations for MOS Prediction}},
  author    = {Wei-Cheng Tseng and Chien-yu Huang and Wei-Tsung Kao and Yist Y. Lin and Hung-yi Lee},
  year      = {2021},
  booktitle = {{Interspeech 2021}},
  pages     = {2781--2785},
  doi       = {10.21437/Interspeech.2021-2013},
  issn      = {2958-1796},
}

@INPROCEEDINGS{cooper2022sslmos,
  author={Cooper, Erica and Huang, Wen-Chin and Toda, Tomoki and Yamagishi, Junichi},
  booktitle={ICASSP 2022 - 2022 IEEE International Conference on Acoustics, Speech and Signal Processing (ICASSP)}, 
  title={Generalization Ability of MOS Prediction Networks}, 
  year={2022},
  volume={},
  number={},
  pages={8442-8446},
  doi={10.1109/ICASSP43922.2022.9746395}}

@article{nakata2024utduss,
  title={UTDUSS: UTokyo-SaruLab System for Interspeech2024 Speech Processing Using Discrete Speech Unit Challenge},
  author={Nakata, Wataru and Yamauchi, Kazuki and Yang, Dong and Hyodo, Hiroaki and Saito, Yuki},
  journal={arXiv preprint arXiv:2403.13720},
  year={2024}
}

@inproceedings{pandey25_ssw,
  title     = {{What is Naturalness?}},
  author    = {Ayushi Pandey and Sébastien {Le Maguer} and Naomi Harte},
  year      = {2025},
  booktitle = {{13th edition of the Speech Synthesis Workshop}},
  pages     = {215--221},
  doi       = {10.21437/SSW.2025-33},
}

@ARTICLE{shirali2026naturalness,
  author={Shirali-Shahreza, Sajad and Penn, Gerald},
  journal={IEEE Open Journal of Signal Processing}, 
  title={Better Naturalness Evaluation of TTS Systems}, 
  year={2026},
  volume={7},
  number={},
  pages={296-304},
  doi={10.1109/OJSP.2026.3657142}}

@inproceedings{dall14_speechprosody,
  title     = {{Rating Naturalness in Speech Synthesis: The Effect of Style and Expectation}},
  author    = {Rasmus Dall and Junichi Yamagishi and Simon King},
  year      = {2014},
  booktitle = {{Speech Prosody 2014}},
  pages     = {1012--1016},
  doi       = {10.21437/SpeechProsody.2014-192},
  issn      = {2333-2042},
}

@inproceedings{kirkland23_ssw,
  title     = {{Stuck in the MOS pit: A critical analysis of MOS test methodology in TTS evaluation}},
  author    = {Ambika Kirkland and Shivam Mehta and Harm Lameris and Gustav Eje Henter and Eva Szekely and Joakim Gustafson},
  year      = {2023},
  booktitle = {{12th ISCA Speech Synthesis Workshop (SSW2023)}},
  pages     = {41--47},
  doi       = {10.21437/SSW.2023-7},
}

@article{huang2026codecmos,
  title={CodecMOS-Accent: A MOS Benchmark of Resynthesized and TTS Speech from Neural Codecs Across English Accents},
  author={Huang, Wen-Chin and Sanders, Nicholas and Cooper, Erica},
  journal={arXiv preprint arXiv:2603.14328},
  year={2026}
}

@article{ren2026mos,
  title={MOS-Bias: From Hidden Gender Bias to Gender-Aware Speech Quality Assessment},
  author={Ren, Wenze and Lin, Yi-Cheng and Huang, Wen-Chin and Cooper, Erica and Zezario, Ryandhimas E and Wang, Hsin-Min and Lee, Hung-yi and Tsao, Yu},
  journal={arXiv preprint arXiv:2603.10723},
  year={2026}
}

@misc{weide1993cmudict,
    title = {The {CMU} Pronouncing Dictionary},
    author = {Weide, Robert L.},
    year = {1993},
    howpublished = {\url{http://www.speech.cs.cmu.edu/cgi-bin/cmudict}} }

@inproceedings{
lee2023bigvgan,
title={Big{VGAN}: A Universal Neural Vocoder with Large-Scale Training},
author={Sang-gil Lee and Wei Ping and Boris Ginsburg and Bryan Catanzaro and Sungroh Yoon},
booktitle={The Eleventh International Conference on Learning Representations },
year={2023},
}

@INPROCEEDINGS{nakata2026sidon,
  author={Nakata, Wataru and Saito, Yuki and Ueda, Yota and Saruwatari, Hiroshi},
  booktitle={ICASSP 2026 - 2026 IEEE International Conference on Acoustics, Speech and Signal Processing (ICASSP)}, 
  title={Sidon: Fast and Robust Open-Source Multilingual Speech Restoration for Large-Scale Dataset Cleansing}, 
  year={2026},
  volume={},
  number={},
  pages={17617-17621},
  doi={10.1109/ICASSP55912.2026.11460553}}

@INPROCEEDINGS{dnsmos,
  author={Reddy, Chandan K A and Gopal, Vishak and Cutler, Ross},
  booktitle={ICASSP 2021 - 2021 IEEE International Conference on Acoustics, Speech and Signal Processing (ICASSP)}, 
  title={Dnsmos: A Non-Intrusive Perceptual Objective Speech Quality Metric to Evaluate Noise Suppressors}, 
  year={2021},
  volume={},
  number={},
  pages={6493-6497},
  doi={10.1109/ICASSP39728.2021.9414878}}

@inproceedings{shirali2023better,
  title={Better replacement for TTS naturalness evaluation},
  author={Shirali-Shahreza, Sajad and Penn, Gerald},
  booktitle={12th Speech Synthesis Workshop (SSW) 2023},
  year={2023}
}

@article{nussbaum2025understanding,
  title={Understanding voice naturalness},
  author={Nussbaum, Christine and Fr{\"u}hholz, Sascha and Schweinberger, Stefan R},
  journal={Trends in Cognitive Sciences},
  volume={29},
  number={5},
  pages={467--480},
  year={2025},
  publisher={Elsevier}
}

@inproceedings{baba2024t05,
  title={The t05 system for the voicemos challenge 2024: Transfer learning from deep image classifier to naturalness mos prediction of high-quality synthetic speech},
  author={Baba, Kaito and Nakata, Wataru and Saito, Yuki and Saruwatari, Hiroshi},
  booktitle={2024 IEEE Spoken Language Technology Workshop (SLT)},
  pages={818--824},
  year={2024},
  _organization={IEEE}
}

@inproceedings{huang25g_interspeech,
  title     = {{SHEET: A Multi-purpose Open-source Speech Human Evaluation Estimation Toolkit}},
  author    = {Wen-Chin Huang and Erica Cooper and Tomoki Toda},
  year      = {2025},
  booktitle = {{Interspeech 2025}},
  pages     = {2355--2359},
  doi       = {10.21437/Interspeech.2025-1977},
  issn      = {2958-1796},
}

@inproceedings{janska10_interspeech,
  title     = {{Native and non-native speaker judgements on the quality of synthesized speech}},
  author    = {Anna C. Janska and Robert A. J. Clark},
  year      = {2010},
  booktitle = {{Interspeech 2010}},
  pages     = {1121--1124},
  doi       = {10.21437/Interspeech.2010-356},
  issn      = {2958-1796},
}

@inproceedings{dong11_blizzard,
  title     = {{I2R Text-to-Speech System for Blizzard Challenge 2011}},
  author    = {Minghui Dong and S W Lee and Paul Chan and Ling Cen},
  year      = {2011},
  booktitle = {{The Blizzard Challenge 2011}},
  pages     = {42--46},
  doi       = {10.21437/Blizzard.2011-8},
}

@inproceedings{chen25b_interspeech,
  title     = {{Fine-Tuning Text-to-Speech Diffusion Models Using Reinforcement Learning with Human Feedback}},
  author    = {Jingyi Chen and Ju Seung Byun and Micha Elsner and Pichao Wang and Andrew Perrault},
  year      = {2025},
  booktitle = {{Interspeech 2025}},
  pages     = {3454--3458},
  doi       = {10.21437/Interspeech.2025-63},
  issn      = {2958-1796},
}

@ARTICLE{seki2025ttsops,
  author={Seki, Kentaro and Takamichi, Shinnosuke and Saeki, Takaaki and Saruwatari, Hiroshi},
  journal={IEEE Transactions on Audio, Speech and Language Processing}, 
  title={TTSOps: A Closed-Loop Corpus Optimization Framework for Training Multi-Speaker TTS Models From Dark Data}, 
  year={2025},
  volume={33},
  number={},
  pages={4956-4970},
  doi={10.1109/TASLPRO.2025.3633089}}

@inproceedings{arora-etal-2026-optimizing,
    title = "Optimizing Conversational Quality in Spoken Dialogue Systems with Reinforcement Learning from {AI} Feedback",
    author = "Arora, Siddhant  and
      Tian, Jinchuan  and
      Shi, Jiatong  and
      Futami, Hayato  and
      Kashiwagi, Yosuke  and
      Tsunoo, Emiru  and
      Watanabe, Shinji",
    editor = "Liakata, Maria  and
      Moreira, Viviane P.  and
      Zhang, Jiajun  and
      Jurgens, David",
    booktitle = "Findings of the {A}ssociation for {C}omputational {L}inguistics: {ACL} 2026",
    month = jul,
    year = "2026",
    address = "San Diego, California, United States",
    publisher = "Association for Computational Linguistics",
    url = "https://aclanthology.org/2026.findings-acl.2040/",
    doi = "10.18653/v1/2026.findings-acl.2040",
    pages = "41049--41066",
    ISBN = "979-8-89176-395-1"
}

@article{Derwing_Munro_1997, 
title={ACCENT, INTELLIGIBILITY, AND COMPREHENSIBILITY: Evidence from FourL1s}, 
volume={19}, 
DOI={10.1017/S0272263197001010}, 
number={1}, 
journal={Studies in Second Language Acquisition}, 
author={Derwing, Tracey M. and Munro, Murray J.}, year={1997}, pages={1–16}
}

@techreport{itu1996p800,
  title        = {Methods for Subjective Determination of Transmission Quality},
  institution  = {International Telecommunication Union},
  type         = {ITU-T Recommendation},
  number       = {P.800},
  year         = {1996}
}

@inproceedings{clark07_blizzard,
  title     = {{Statistical analysis of the Blizzard Challenge 2007 listening test results}},
  author    = {Robert A. J. Clark and Monika Podsiadlo and Mark Fraser and Catherine Mayo and Simon King},
  year      = {2007},
  booktitle = {{Blizzard 2007}},
  _pages     = {paper 003},
}

@article{lertpetchpun2026learning,
  title={Learning-free L2-Accented Speech Generation using Phonological Rules},
  author={Lertpetchpun, Thanathai and Lee, Yoonjeong and Lee, Jihwan and Feng, Tiantian and Byrd, Dani and Narayanan, Shrikanth},
  journal={arXiv preprint arXiv:2603.07550},
  year={2026}
}

\end{document}